\documentclass{emulateapj}

\usepackage{amssymb}
\usepackage{graphicx}

\slugcomment{Accepted by ApJ, March 9th, 2007}
\shorttitle{Spitzer view of I Zw 18}
\shortauthors{Wu et al.}

\newcommand{\degree}{\ensuremath{^\circ}}

\begin{document}

\title{Dust in the Extremely Metal-Poor Blue Compact Dwarf Galaxy
  I\,Zw\,18: the Spitzer Mid-infrared View}

\author{Yanling Wu\altaffilmark{1}, V. Charmandaris\altaffilmark{2,3},
  L. K.~Hunt\altaffilmark{4}, J. Bernard-Salas\altaffilmark{1},
  B. R.~Brandl\altaffilmark{5}, J. A.~ Marshall\altaffilmark{1},
  Vianney Lebouteiller\altaffilmark{1}, Lei Hao\altaffilmark{1},
  J. R.~Houck\altaffilmark{1}}

\altaffiltext{1}{Astronomy Department, Cornell University, Ithaca, NY 14853}

\altaffiltext{2}{University of Crete, Department of Physics, GR-71003,
Heraklion, Greece}

\altaffiltext{3}{IESL/Foundation for Research and Technology - Hellas,
  GR-71110, Heraklion, Greece, and Chercheur Associ\'e, Observatoire de
  Paris, F-75014, Paris, France}

\altaffiltext{4}{INAF-Istituto di Radioastronomia, Largo E. Fermi 5,
50125 Florence, Italy}

\altaffiltext{5}{Leiden Observatory, Leiden University, P.O. Box 9513,
2300 RA Leiden, The Netherlands}

\email{wyl@astro.cornell.edu, vassilis@physics.uoc.gr,
  hunt@arcetri.astro.it, jbs@isc.astro.cornell.edu,
  brandl@strw.leidenuniv.nl, jam258@cornell.edu,
  vianney@isc.astro.cornell.edu,haol@astro.cornell.edu,
  jrh13@cornell.edu}

\begin{abstract}
I\,Zw\,18, a blue compact dwarf (BCD) galaxy with the 2nd lowest
metallicity measured in a star-forming object, has been observed with
all three instruments on board the Spitzer Space Telescope. We present
the deepest 5$-$36\,$\mu$m mid-infrared (mid-IR) spectrum of this
galaxy as yet obtained, as well as 3.6\,$\mu$m to 70\,$\mu$m imaging
results. As with SBS\,0335-052E, another BCD with similar metallicity,
I\,Zw\,18 shows no detectable emission from polycyclic aromatic
hydrocarbons (PAHs). However, the continuum emission, from 15 to
70\,$\mu$m, of I\,Zw\,18 has a much steeper slope, more characteristic
of a typical starburst galaxy of solar abundance.  The neon abundance
as measured from the infrared fine-structure lines is $\sim$
1/23\,Z$_\odot$, and the sulfur abundance is $\sim$ 1/35\,Z$_\odot$,
generally consistent with the nebular oxygen abundance of
1/30\,Z$_\odot$ derived from optical lines. This suggests that the
extinction to the infrared emitting regions of this galaxy is low, 
also in agreement with the optical Balmer line ratios.

\end{abstract}


\section{Introduction}

Low-metallicity galaxies may have been the first sites of star
formation in the early universe \citep{White78, Dekel86}. Consequently
understanding their properties may provide valuable insights into the
formation of the first generation of normal stars and the subsequent
enrichment of the interstellar medium.  However, finding truly
primordial galaxies in the distant universe is beyond the reach of current
technology. At high redshift, observational limitations introduce
biases toward the detection of high-mass high-luminosity systems in
which the short time-scales of massive star formation lead to the
identification of systems that are already chemically enriched
\citep[i.e.,][]{Maiolino03}.  An alternative
approach is to identify and study unevolved galaxies in the local
universe. Such a sample are the blue compact dwarf galaxies
\citep{Kunth00}.

Since its discovery by \citet{Zwicky66} and the seminal paper of
\citet{Searle72}, I\,Zw\,18 has been studied extensively at many
wavelengths. With an oxygen abundance determined from the
optical lines in H{\sc ii} regions of 12+log(O/H)=7.17
\citep{Skillman93,Izotov99}, or
$\sim$1/30\,Z$_\odot$\footnote{\citet{Izotov99} give 1/50\,Z$_\odot$.
Here, we use the new oxygen solar abundance of 12+log[O/H]=8.69
\citep{Prieto01}, which results in a metallicity of 1/30\,Z$_\odot$.},
it has remained the lowest metallicity BCD for over two decades until
the recent study of the western component of SBS\,0335-052
\citep{Izotov05,Papaderos06}, which has slightly lower metallicity
(12+log(O/H)=7.12). Distance estimates to I\,Zw\,18 range from
$\sim$10 Mpc \citep{Hunter95}, $\sim$12.6 Mpc \citep{Ostlin00}, and up 
to $\sim$15 Mpc \citep{Izotov04}. Here we adopt a distance of 12.6 Mpc 
(1$\arcsec$$\sim$61\,pc).  
I\,Zw\,18 consists of two bright knots of
star-formation, a northwest (NW) component and a southeast (SE) one,
together they form the ``main body'' of the system. Both the NW and SE
components contain numerous young star clusters, with ages ranging
from 3 to 10\,Myr, but the age of the underlying stellar population in
I\,Zw\,18 is still a matter of debate, with maximum ages ranging from
500\,Myr to 5\,Gyr \citep{Aloisi99,Ostlin00,
Recchi02,Hunt03,Izotov04,Momany05}. Approximately 22$\arcsec$
northwest of the ``main body'', there is a blue irregular low surface
brightness star-forming region, called the ``C component'', which is
embedded in a common H{\sc i} envelope \citep{vanZee98,Izotov04}.

It is well known that some BCDs are forming stars at a rate that can
only be maintained for $\sim$1/10 of Hubble time given their store of
available hydrogen. SBS\,0335-052E, which has a similar metallicity
(12+log[O/H]=7.31), is forming stars at a high rate in dense compact
regions \citep{Houck04b}.  I\,Zw\,18 is forming stars at a slower rate
in complexes which are diffuse and extended \citep{Hunt05a}. These
differences have been attributed to different star-formation
modes. SBS\,0335-052E is in an ``active'' mode with relatively high
star-formation rate (SFR) in compact dense regions, while in I\,Zw\,18
star-formation occurs in a ``passive'' mode in more extended regions
with a relatively low SFR \citep{Hirashita04}.

In this paper, we present a detailed analysis of the infrared
properties of the ``main body'' of I\,Zw\,18 based on the current
deepest mid-IR spectra for this galaxy, obtained using the infrared
spectrograph \citep[IRS,][]{Houck04a} on the Spitzer Space Telescope
\citep{Werner04}. We also discuss the results of photometric
observations with IRAC \citep{Fazio04} and MIPS \citep{Rieke04}
observations. We describe the observation and data analysis in \S
2. The Spitzer images and spectra are discussed in \S 3, along with
the observed morphologies and abundance estimates.  A discussion of
the SFR of I\,Zw\,18 is given in \S 4 and we summarize our conclusions
in \S 5.

\section{Observations}

\subsection{Spitzer/IRS Spectroscopy}

I\,Zw\,18 was observed as part of the IRS\footnote{The IRS was a
collaborative venture between Cornell University and Ball Aerospace
Corporation funded by NASA through the Jet Propulsion Laboratory and
the Ames Research Center.} Guaranteed Time Observation program
on 27 March 2004 using all four instrument modules. It was re-observed
on 23 April 2005 with the short-low (SL) and long-low (LL) modules
with increased integration time to achieve a higher signal-to-noise
ratio (SNR) over the $5-36\,\mu$m mid-IR continuum. A third even
longer observation was made on 16 December 2005 with all four IRS modules. 
The target was acquired using the 22$\mu$m (red) peak-up
camera in high-accuracy mode, and the details of the observations are 
presented in Table \ref {tab1}.

The data were processed at the {\em Spitzer} Science Center (SSC) 
(pipeline version 14.0). The two-dimensional image data were
converted to slope images after linearization correction, subtraction
of darks, and cosmic-ray removal. Finally, the data were co-added. \
In order to increase the SNR of the subtracted background for SL and
LL, we combined the background observed in off-order and off-nod
positions.  A detailed explanation for this method on faint source
extraction can be found in \citet{Weedman2006}. After subtracting the
background, the one-dimensional spectra were extracted from images 
with a script version of the Spectral Modeling, Analysis and Reduction Tool 
(SMART, ver. 6.0.4, \citet{Higdon2004}). We used tapered column extraction 
starting from intermediate pipeline product droop files, which only lack 
stray light and flat-field correction. The data from short-high (SH) and 
long-high (LH) droop files used the full slit extraction method
from the mean of the combined images. We calibrated the flux densities
by multiplying the extracted spectrum with the relative spectral
response function (RSRF), which was created from the IRS
standard stars, HR6348 for SL, HD173511 for LL and $\xi$ Dra for SH
and LH, for which accurate templates are available \citep{Cohen03}.

\subsection{Spitzer Imaging with IRAC and MIPS}

The galaxy was imaged at 3.6, 4.5, 5.8 and 8\,$\mu$m using IRAC on 
3 April 2004, as well as at 24, 70 and 160\,$\mu$m using MIPS on 7 April 
2004 \citep{Engelbracht05} (PID:59) (See Table \ref{tab1}).  The
IRAC high dynamic range mode was used with a 4 point small
cycling pattern of 30 seconds exposure time for each frame. This
resulted in an on-source time of 120 seconds for each IRAC filter. The
MIPS photometry mode was used in small
fields with 1 cycle $\times$ 3 seconds at 24\,$\mu$m and 2 cycles
$\times$ 10 seconds at both 70 and 160\,$\mu$m. 
Two offset positions ($\pm12\arcsec$) were used to allow proper
subtraction of bad pixels. The total on-source times were 48, 231 and
42 seconds for the MIPS 24, 70 and 160\,$\mu$m bands. The
imaging data were processed by the SSC pipeline version 14.0 and the
final mosaics were obtained from the Spitzer archive.

\begin{deluxetable}{cclr}
\tabletypesize{\scriptsize}
\setlength{\tabcolsep}{0.02in}
\tablecaption{Spizer Observations of I\,Zw\,18\label{tab1}}
\tablewidth{0pc}
\tablehead{ 
  \colhead{AORKEY}  & \colhead{Date} & \colhead{Instrument} & \colhead{On-source Time} \\
    \colhead{} & \colhead{} & \colhead{} & \colhead{(sec)} \\
}
\startdata
9008640  & 2004-03-27 & IRS (SL) &  84 \\
         &            & IRS (LL) & 120 \\
         &            & IRS (SH) & 240 \\
         &            & IRS (LH) & 120 \\
4330759  & 2004-04-03 & IRAC (3.6\,$\mu$m) & 120 \\
         &            & IRAC (4.5\,$\mu$m) & 120 \\
         &            & IRAC (5.8\,$\mu$m) & 120 \\
         &            & IRAC (8.0\,$\mu$m) & 120 \\
4349184  & 2004-04-07 & MIPS (24\,$\mu$m)  &  48 \\
         &            & MIPS (70\,$\mu$m)  & 231 \\
         &            & MIPS (160\,$\mu$m) &  42 \\
12622848 & 2005-04-23 & IRS (SL) & 480 \\
         &            & IRS (LL) & 240 \\
16205568 & 2005-12-16 & IRS (SL) &2040 \\
         &            & IRS (LL) & 840 \\
         &            & IRS (SH) &2880 \\
         &            & IRS (LH) &1440 \\
\enddata

\end{deluxetable}

\section{Results}

\subsection{Mid-IR Morphology}

\begin{figure*}
  \epsscale{1.4}
  \plotone{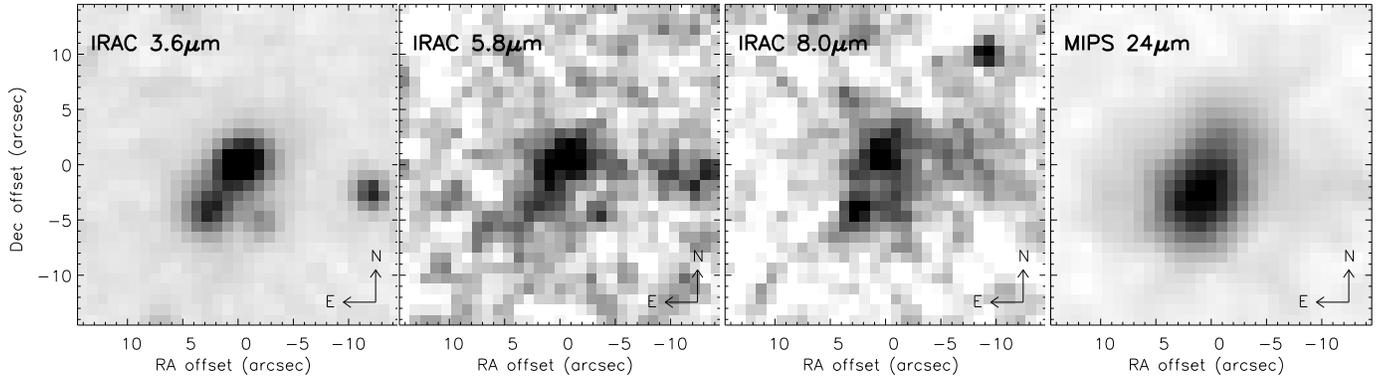}
  \caption{IRAC 3.6, 5.8, 8\,$\mu$m and MIPS 24\,$\mu$m images of
    I\,Zw\,18 as observed by {\em Spitzer}. The galaxy is resolved into
    a NW and a SE component at 3.6\,$\mu$m which gradually become
    blended to a single extended source at 24\,$\mu$m. At the distance
    of 12.6 Mpc, 1$\arcsec$ corresponds to $\sim$61 pc. The image is
    centered on the optical centroid of the galaxy
    ($\alpha=09\degree34\arcmin02.0\arcsec$,
    $\delta=55\degree14\arcmin28\arcsec$). Note the change in
    morphology and peak of the emission as the wavelength increases.}
  \label{fig:fig1}
\end{figure*}
Many ground-based and space-born instruments have been used to
obtain optical/UV to near-IR data for I\,Zw\,18 \citep{Hunter95,
Aloisi99, Ostlin00, Cannon02, Hunt03, Izotov04}. \citet{Hunter95}
first resolved the ``main body'' of I\,Zw\,18 into stars using the
{\em Hubble} Space Telescope (HST). They have also detected filaments of
ionized gas up to 450\,pc from the center of the galaxy. Keck II
spectra revealed H$\alpha$ emission as far as $\sim$1800\,pc from the
``main body'' of I\,Zw\,18. \citet{Izotov01} have also shown that the
equivalent widths of emission lines are large in this extended envelope.
This, together with the optical and near-IR colors, suggests that 
ionized gas dominates the emission in the outermost regions.

In Figure 1 we present images of the ``mainbody'' of I\,Zw\,18 in four
infrared bands (3.6, 4.5, 8 and 24\,$\mu$m).  At 3.6\,$\mu$m, where
most of the light is due to the stellar photospheric emission, the
morphology of I\,Zw\,18 is very similar to that in deep near-IR
imaging \citep{Hunt03} and broad band optical optical imaging
\citep{Izotov04}. The NW component is noticeably more extended and
brighter than the SE one. At 8$\mu$m the components are still clearly
resolved while the contrast in the brightness between the two
components has decreased.  In normal star forming galaxies the
emission sampled by the IRAC 8\,$\mu$m filter is typically dominated
by dust continuum, and PAH emission when PAHs are present. Some
continuum emission from the nearly Rayleigh-Jeans tail of stellar
photospheric emission may also be present, even though its
contribution is typically small in late type or irregular galaxies
\citep[see][]{Smith07}. As we discuss in the following section, no PAH
features are detected in the IRS spectrum of I\,Zw\,18 down to our
1\,$\sigma\sim$0.2\,mJy sensitivity limit. To estimate the
contribution of the stellar continuum to the observed 8\,$\mu$m flux
density we follow the approach of \citet{Jackson06} and apply a scale
factor of 0.4 in the 4.5\,$\mu$m emission from the galaxy. This
suggests that no more than $\sim$25\% of the 8\,$\mu$m flux can be
attributed to the stars \citep{Engelbracht05}. This was also to be
expected given the observed steeply rising slope of the mid-IR spectrum. 
Therefore most of the ``mainbody''
emission seen in the 8$\mu$m band is due to dust continuum emission.
Hence we interpret this gradual shift in brightness from the NW to the
SE component to the probable presence of more embedded star formation
in the SE component which was obscured in the optical broad band
imaging. Interestingly, while optical recombination line ratios
give an average extinction of only A$_{\rm V}$$\sim$ 0.2\,mag, there
are also statistically significantly high H$\alpha$/H$\beta$ flux
ratios ($\sim$3.4, corresponding to A$_{\rm V}$$\sim$ 0.5\,mag) in the
SE component \citep{Cannon02}, suggesting the existence of an
appreciable amount of dust within the galaxy.  Moreover, as can be
seen in Figure 2, there is some extended 8$\mu$m emission, though at
low levels, to the west of the NW component. This emission has similar
morphology as the radio continuum emission detected in the X-band and
L-band by \citet{Cannon05} and \citet{Hunt05b}, which has been
attributed to low-frequency flux from a synchrotron halo.

\begin{figure}
  \epsscale{1.0}
  \plotone{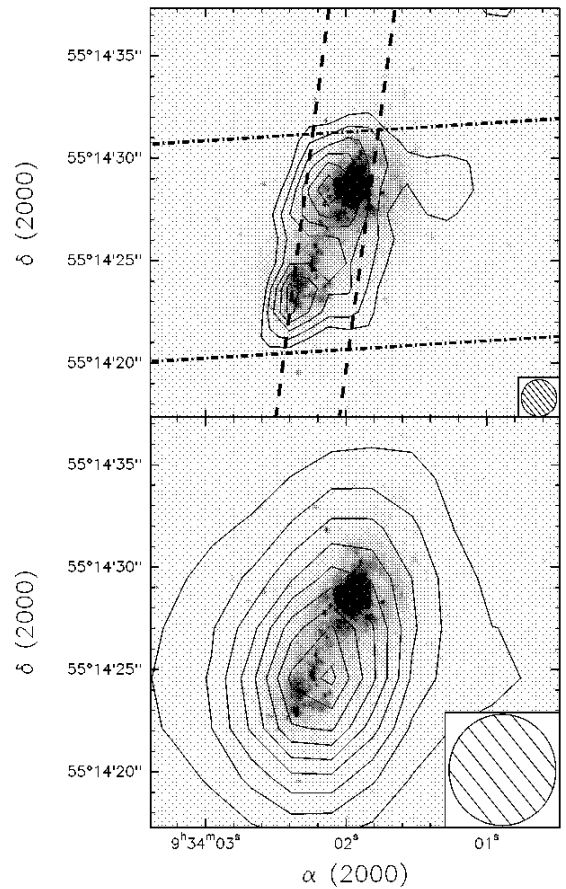}
  \caption{a) Top: Contour overlay of the IRAC 8\,$\mu$m image on the
    {\em HST}/WFPC2 F555W image taken from the HST archive. The
    astrometric calibration of the HST image has been derived using
    stars from the U.S. Naval Observatory Astrometric Catalog B1.0,
    and is the same image as used in \citet{Hunt05b}.  The contours
    range from 4$\sigma$ ($\sigma$=0.03 MJy\,sr$^{-1}$) above the sky
    level (1.40 MJy\,sr$^{-1}$) to the peak value of 1.71
    MJy\,sr$^{-1}$. The location of the IRS SL (3.6$\arcsec$ in width) 
    and LL (10.5$\arcsec$) slits, are also indicated with the dashed and 
    dash-dotted lines respectively. Note that due to the extent of the
    galaxy, and the fixed position angle of the slit, part of both the
    NW and SE components are not fully covered by the SL slit. The size of 
    the PSF at 8\,$\mu$m (1.8$\arcsec$) is also shown at the lower right 
    of the panel. b) Bottom: the same optical image with the contours of MIPS
    24\,$\mu$m emission (from 4$\sigma$ and above). The size of the PSF is 
    5.4$\arcsec$.}
\label{fig:fig2}
\end{figure}

In Figure 2, we show contour overlays of the 8\,$\mu$m and 24\,$\mu$m
emission on the {\em HST} V-band image of the ``main body'' of
I\,Zw\,18. At 8\,$\mu$m, the source is clearly resolved into two components. 
The centroids of the two components, especially the NW one, are
slightly displaced from their optical counterparts.
Moreover, this displacement becomes even more pronounced at longer wavelengths.
At 24\,$\mu$m, the two components are blended into a single source, the centroid 
of which is located slightly closer to the SE region. This displacement is real
and indicates the presence of more 24\,$\mu$m dust emission in the SE
cluster. To confirm this change in morphology we convolved the
8\,$\mu$m image to the $\sim5.4\arcsec$ size of the
24\,$\mu$m point-spread-function (PSF). Even though the resulting
marginally resolved source is also elongated in the SE to NW direction,
the peak emission was further to the NW than the peak of the 24$\mu$m image. 
This suggests an actual change in the spatial distribution of the various 
dust temperature components in the galaxy. In
Figure 2, we also overlay the IRS SL/LL slit on the image of the
galaxy. As one can see, since the SL slit is only 3.6$\arcsec$ wide,
some of the flux from both components is not properly sampled,
resulting in an underestimate of the extended emission from the
galaxy. However, because the spectrograph is not sensitive
to the low-surface brightness emission from the areas denoted by 
the lowest 8\,$\mu$m contours, this does not bias our analysis of the global spectral 
properties of the system or affects any of the conclusions drawn.

\subsection{Mid-IR Spectral Features}

Figure 3 shows the 5.3-36\,$\mu$m low-resolution spectrum of I\,Zw\,18
as observed by the IRS.  The SNR is $\sim$3 times higher than that
shown by \citet{Wu06}.  As we will discuss in \S 3.4, the global
shape of the mid-IR spectrum, reveals that the IRS spectrum of
I\,Zw\,18 continues to rise steeply with wavelength from 5\,$\mu$m all
the way to 36\,$\mu$m. This is unlike the case of SBS\,0335-052E
\citep{Houck04b}, the third lowest metallicity galaxy to date, which
has a nearly flat continuum peaking at $\sim$28\,$\mu$m in f$_\nu$.

The improved SNR of the new spectrum enables us to detect for the 
first time several mid-IR forbidden lines. 
Fine structure lines, such as [SIV] at 10.51\,$\mu$m and [NeIII] at
15.55\,$\mu$m can clearly be seen, even in the low-resolution
spectrum. Several additional forbidden lines, such as [NeII] at
12.81\,$\mu$m, [SIII] at 18.71 and 33.48\,$\mu$m, as well as [SiII] at
34.82\,$\mu$m are identified in the high-resolution spectrum (see
Figure 4). [OIV] at 25.89\,$\mu$m and [FeII] at 25.99\,$\mu$m are blended 
in LL, but can clearly be resolved in LH. The line
fluxes measured from the IRS high-resolution spectrum are reported in
Table \ref{tab2}.  The observed line ratio of
[NeIII]/[NeII] is $\sim$ 5 and the ratio
of [SIV]/[SIII](18.71\,$\mu$m) is $\sim$ 2. This
indicates that the radiation field in I\,Zw\,18 is much harder than in
a typical starburst galaxy, where [NeIII]/[NeII] is usually $\le$ 1
\citep{Brandl06}, and even harder than the majority of the BCDs
observed so far \citep{Hunt06, Wu06}.

\begin{figure}
  \epsscale{1.2}
  \plotone{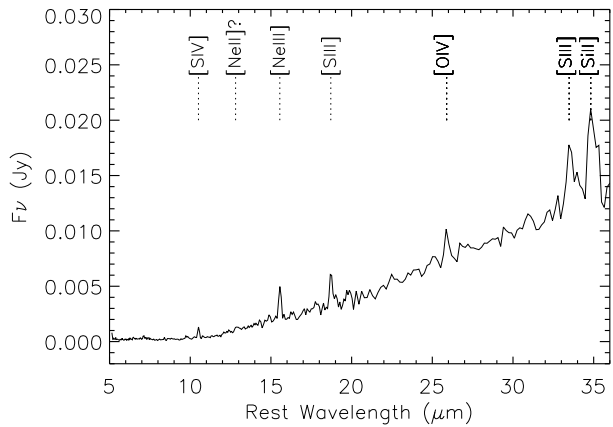}
  \caption{{\em Spitzer/IRS} 5-36\,$\mu$m low-resolution spectrum of
    I\,Zw\,18.  No scaling factors have been applied to stitch the
    different orders and modules. We indicate several of the well
    known mid-IR fine structure emission lines, detected in the
    spectrum (see also Fig. 4).}
  \label{fig:fig3}
\end{figure}

\begin{figure}
  \epsscale{1.3}
  \plotone{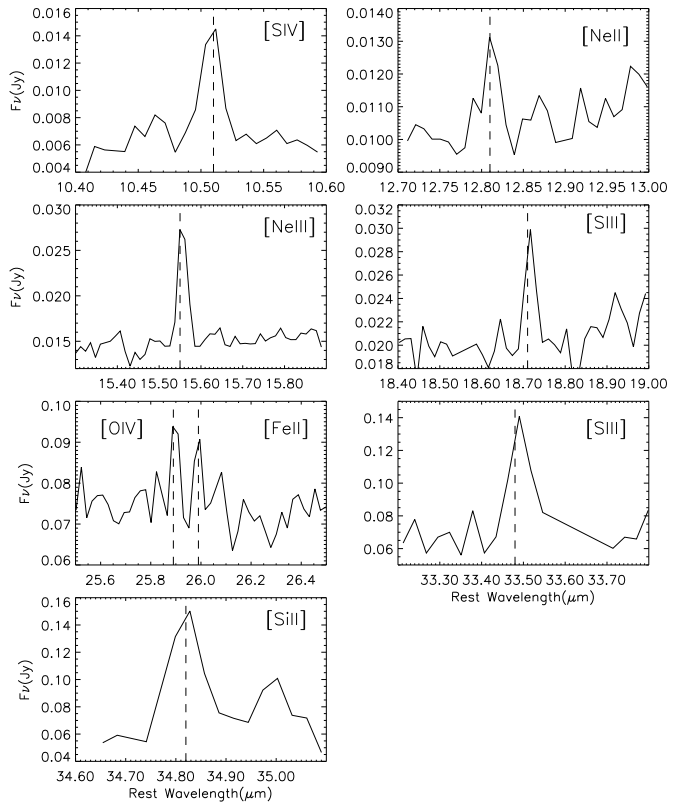}
  \caption{Mid-IR fine structure lines of [SIV] (10.51$\mu$m), [NeII]
  (12.81$\mu$m), [NeIII] (15.55$\mu$m), [SIII](18.71, 33.42$\mu$m)
  [OIV] (25.89$\mu$m), [FeII] (25.99$\mu$m) and [SiII] (34.82$\mu$m)
  from the high-resolution spectrum of I\,Zw\,18. Note that the sky
  emission has not been subtracted.}
  \label{fig:fig4}
\end{figure}

As was the case with SBS\,0335-052E, the 5-15\,$\mu$m spectrum of
I\,Zw\,18 does not show any detectable Polycyclic Aromatic Hydrocarbon (PAH)
emission. Using the new mid-IR spectrum, we measure a 3$\sigma$ upper limit of
3.5$\times$10$^{-22}$W{\rm cm}$^{-2}$ for the 6.2\,$\mu$m PAH feature, as
well as an equivalent width (EW) of $<$0.23\,$\mu$m.
The 11.2\,$\mu$m PAH emission has an
upper limit of 1.3$\times$10$^{-22}$W{\rm cm}$^{-2}$ and EW of
$<$0.12\,$\mu$m. The decreasing strength of the PAH features in low
metallicity environment is a current topic of interest
\citep[see][]{Engelbracht05, Wu06,OHalloran06,Jackson06,Madden06, Beirao06}. 
The exact reason why PAHs tend to be absent in
low-metallicity, high-excitation star-forming regions is not yet
clear, although it almost certainly has to do with a combination of
several effects, such as low carbon abundance, shock destruction of 
grains by supernovae, high ionization or excitation resulting from 
low metallicity and/or extreme radiation field intensity.

\subsection{Neon and Sulfur Abundances}

The fine-structure lines of sulfur and neon measured in the
high-resolution spectrum can be used to derive the ionic abundances
relative to hydrogen. To derive these abundances one needs at
least one hydrogen recombination line (usually H$\beta$), as well as 
an estimate of the electron temperature (T$_e$) and electron density (N$_e$).
The H$\beta$ fluxes for I\,Zw\,18 reported in the literature often correspond to small
regions of the galaxy or of one of the two components. The IRS SH slit 
from which the IR lines were measured contains part of both components.
We derive the H$\beta$ flux from the thermal component of the 3.6\,cm 
continuum \citep{Cannon05}, which is not affected by extinction effects 
and encompasses the entire ``main body''. The SH slit includes $\sim$52\% 
of the total H$\beta$ emission and this gives an H$\beta$ flux of 
6.1$\times10^{-14}$ergs\,cm$^{-2}$s$^{-1}$. The reported electron
temperatures in the literature range from 18\,000 to 20\,000~K for
\ion{O}{2} and \ion{O}{3} \citep{Skillman93, Izotov99, Thuan05,
Shi05}, with a slightly lower value for \ion{S}{2} (T$_e\sim$15\,500
K). The infrared lines are less sensitive to T$_e$ and here 
we adopt an average temperature of 19\,000\,K. An electron density 
(N$_e$) of 100~cm$^{-3}$ was assumed \citep{Shi05,Izotov99}. 
Using the above quantities and
resolving the population of the levels for each ion, the ionic
abundances can be derived (see \cite{Bernard-Salas01}, their eq. 1).

The calculated ionic abundances are presented in Table \ref{tab2}. The
elemental abundances can be obtained by adding the contribution of
each ion. In the case of neon, only \ion{Ne}{2} and \ion{Ne}{3} are 
observed. The presence of \ion{O}{4} suggests that some \ion{Ne}{4} may 
be present. Assuming that the contribution of \ion{Ne}{4}
is 10\% of that of \ion{Ne}{3} as has been found in some planetary
nebulae (PN\footnote{The high excitation and densities in BCDs are a
closer match to PN than the typical H~{\sc ii} regions found in normal
starburst galaxies.}, \cite{Bernard-Salas03}), the total neon
abundance is 5.3$\times$10$^{-6}$, which is 1/23 Z$_\odot$. 
Similarly, adding up the ionic abundance of \ion{S}{3} and \ion{S}{4},
the total sulfur abundance is 4.0$\times$10$^{-7}$, which is 1/35
Z$_\odot$. However, this might be a lower limit because sulfur abundance 
as derived from galactic planetary nebulae and H{\sc ii} regions is lower 
than the solar sulfur abundance \citep{Pottasch06,Maciel03}, probably 
due to an overestimate of the solar value. Comparing directly with the ionic 
abundance from the optical \citep{Izotov99}, our \ion{Ne}{3} abundance is 
nearly twice as high while \ion{S}{3} abundance agrees quite well. The 
difference in \ion{Ne}{3} could come from the electron temperature that 
is used to derived the ionic abundance. The optical is more affected 
by the change in temperature as compared to the infrared. 
Lowering T$_e$ from 19000\,K to 15000\,K would double the \ion{Ne}{3} abundance 
derived from the 3869$\AA$ line. It is known that in some PN, the temperature 
obtained from the \ion{Ne}{3} ion is lower than that from the \ion{O}{3} 
\citep{Bernard-Salas02}. Another possible explanation is that there are
some regions with dust obscuring the optical emission lines.  Overall, 
the neon (1/23\,Z$_\odot$) and sulfur (1/35\,Z$_\odot$) abundances we derive 
using the infrared lines are consistent with the nebular oxygen abundance 
(1/30\,Z$_\odot$), which supports the low extinction (A$_{\rm V}$=0.2\,mag) 
derived from hydrogen recombination lines  by \citet{Cannon02}.

\begin{deluxetable}{c c c c}
\tablewidth{8cm} \tablecaption{Fluxes and ionic
abundances\tablenotemark{a} \label{tab2}} 

      \tablehead{\colhead{$\lambda_{\rm rest}$} & \colhead{Feature} & \colhead{Obs.
Flux\tablenotemark{b}} &
      \colhead{$N_{\rm ion}$/$N_{\rm p}$\tablenotemark{c}}}

\startdata
10.51                     & \ion{[S}{4]}   &  4.8$\pm0.3$     &   1.3$\times$10$^{-7}$  \\
12.81                     & \ion{[Ne}{2]}  &  0.9$\pm0.1$     &   1.4$\times$10$^{-6}$  \\
15.55                     & \ion{[Ne}{3]}  &  4.6$\pm0.2$     &   3.6$\times$10$^{-6}$  \\
18.71\tablenotemark{d}    & \ion{[S}{3]}   &  2.3$\pm0.2$     &   2.8$\times$10$^{-7}$  \\
25.89                     & \ion{[O}{4]}   &  4.9$\pm0.3$     &   \nodata \\
25.99                     & \ion{[Fe}{2]}  &  3.4$\pm0.3$     &   \nodata \\
33.48\tablenotemark{d}    & \ion{[S}{3]}   &  12.0$\pm1.2$    &   \nodata  \\
34.82                     & \ion{[Si}{2]}  &  15.8$\pm1.8$    &   \nodata \\
\enddata

\tablenotetext{a}{Only the abundances of ions used to calculate the metallicities 
  in this work are listed in this table.}
\tablenotetext{b}{In units of 10$^{-15}$~erg~cm$^{-2}$s$^{-1}$.}
\tablenotetext{c}{Ionic abundance relative to hydrogen.}
\tablenotetext{d}{Here we use the \ion{[S}{3]} at 18.71\,$\mu$m when calculating the 
  sulfur abundance because this line is in the SH module and provides us a more 
  reliable measurement as compared to the 33.48\,$\mu$m line which is at the edge of the 
  LH module and is noisy.}

\end{deluxetable}

\subsection{Comparison with SBS\,0335-052E and NGC7714}

I\,Zw\,18 and SBS\,0335-052E share some properties but are very
different in other aspects. Perhaps the most salient difference
between their spectral energy distribution (SED) is the fraction of
their luminosities emitted in the IR. Using the 15\,$\mu$m and
30\,$\mu$m flux densities of I\,Zw\,18 and applying an empirical
relation in starburst galaxies \citep{Brandl06}, we derive L$_{\rm
IR}$=1.8$\times$10$^7$\,M$_\odot$.  While SBS\,0335-052E has L$_{\rm
IR}\sim$ 10$^9$ L$_\odot$ and L$_{\rm IR}$/L$_B\sim$ 1.3, I\,Zw\,18
has L$_{\rm IR}\sim$ 10$^7$ L$_\odot$ and L$_{\rm IR}$/L$_B\sim$ 0.3;
the relative infrared luminosity is a factor of 4 times greater in
SBS\,0335-052E.  Another important difference in the infrared SEDs of
I\,Zw\,18 and SBS\,0335-052E is the peak wavelength of the SED.  The
SED of SBS\,0335-052E peaks at $\sim$28\,$\mu$m in $f_{\nu}$ space
\citep{Houck04b}, indicating very little cold dust.  I\,Zw\,18 has a
clear detection of 34$\pm2.4$\,mJy at 70\,$\mu$m(Engelbracht et
al. 2007, in preparation), and the ratio of f$_{70}$/f$_{24}$ is more
than a factor of 5 while the same ratio in SBS\,0335-052 is less than
1. This suggests that contrary to SBS\,0335-052E, which is a similarly
low metallicity BCD, there is a significant amount of cold dust in
I\,Zw\,18.

We can further compare the properties of I\,Zw\,18 with
higher-luminosity and more metal-rich starbursts.  In Figure 5, we
present the spectra of I\,Zw\,18 and SBS\,0335-052E, together with a
typical starburst galaxy, NGC7714 \citep{Brandl04}, all of which have
been normalized to the flux density of I\,Zw\,18 at 22\,$\mu$m.
Setting aside the strong PAH emission features present in
NGC7714, there is a striking similarity between the mid-IR continuum
of I\,Zw\,18 and NGC7714, although the latter has a metallicity of
more than half solar.  At short mid-IR wavelengths ($\lambda
<$10$\mu$m), the warm dust component dominates and it does not change
much even for an extended starburst galaxy \citep{Brandl06}. At longer
wavelengths, emission from larger cooler grains is dominant.
After normalization at 22\,$\mu$m, the 70\,$\mu$m flux of I\,Zw\,18
differs less than 20\% when compared to the 70\,$\mu$m flux of 9.53 Jy
(Engelbracht, private communication) or the {\em IRAS} 60\,$\mu$m flux
of 11.16 Jy for NGC7714.  Figure 5 suggests that similarly low
metallicity galaxies can have both very flat or very steep spectral
slope, while in the latter case, the spectral slope can be as steep as
that of a typical starburst.  This leads us to conclude that
metallicity is not the main parameter driving the difference in the
shape of the mid-IR spectral slope \citep{Wu06} and the infrared part
of the SEDs.

\begin{figure}
  \epsscale{1.2}
  \plotone{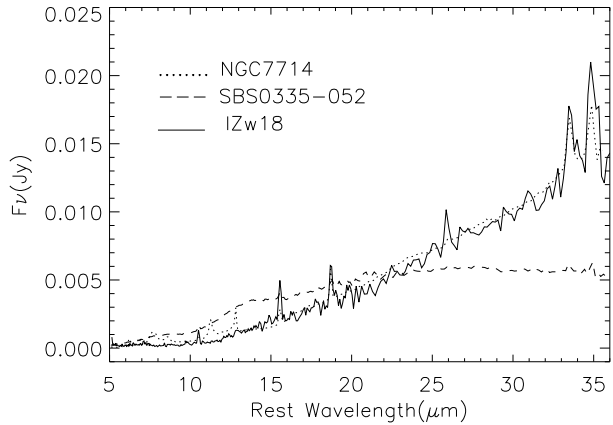}
  \caption{The mid-IR spectra of I\,Zw\,18, SBS\,0335-052E and NGC7714
    between 5-36\,$\mu$m.  The spectra have been normalized to the
    flux density of I\,Zw\,18 at 22\,$\mu$m. SBS\,0335-052E stands out as
    having a very flat continuum in the mid-IR, indicating there is
    much less cold dust in this galaxy. I\,Zw\,18 and NGC7714 display
    a striking resemblance in this wavelength range.}
  \label{fig:fig6}
\end{figure}

Based on the new Spitzer measurements of the mid- and far-infrared
emission from I\,Zw\,18 one could in principle attempt to model the global
SED of the galaxy. We did explore this avenue using modelling tools 
such as DUSTY \citep{Ivezic99}, but with only limited
success. The complex geometry of IZw18 and the large number of free
parameters in the available models prevented us from significantly 
constraining the physical conditions of the dust. We thus refrain from 
elaborating on these results until more data are available.

\section{Star Formation Rate in I\,Zw\,18}

Deriving the star formation rate in nearby galaxies from various
observational indicators and understanding the possible variations in
the results is extremely useful for ``predicting'' the properties of
high-redshift galaxy populations, where only sparse data are
available. Using the available data, we calculated the SFRs of
I\,Zw\,18 from different indicators and we present our results in
Table \ref{tab3}.

The SFR estimated from the H$\alpha$ luminosity gives a value of 
0.05\,M$_\odot$yr$^{-1}$ \citep{Kennicutt94, Cannon02}. However, I\,Zw\,18 
has only 1/30\,Z$_\odot$ and lower metallicities may result in a reduced 
SFR for a given H$\alpha$ luminosity \citep{Lee02,Rosenberg06}. Using the 
metallicity correction recipe of \citet{Lee02}, the SFR would be 
$\sim$0.03\,M$_\odot$yr$^{-1}$.

The radio continuum emission is another important diagnostic of
star-formation process and it is not affected by dust extinction
effects. The thermal free-free emission is a direct indicator of SFR
and it is typically only $\sim$10\% of the total radio continuum at
1.4\,GHz for normal galaxies \citep{Condon92}. However, in I\,Zw\,18,
the fraction of thermal component is three times the typical value
($\sim$30\%: \citealt{Hunt05b,Cannon05}). Using the relation
between radio free-free emission and SFR, we derive a
SFR=0.1\,M$_\odot$yr$^{-1}$ \citep{Hunt05b}.  The non-thermal
component of the radio continuum can also be used to calculate SFR and
we derive a SFR of 0.03\,M$_\odot$yr$^{-1}$ \citep{Condon92}, a factor
of 3 lower than the thermally-derived SFR. This difference is a direct consequence
of the unusual value of the ``thermal/non-thermal'' fraction in
I\,Zw\,18. This is an important caveat that should be considered when
applying the standard correlations, which have been established for normal
star forming galaxies, in very young low-metallicity systems.

Finally, the SFRs estimated from the infrared are significantly lower.
Using the total infrared luminosity of 1.8$\times$10$^7$\,L$_\odot$,
we derive a SFR=0.003\,M$_\odot$yr$^{-1}$ \citep{Kennicutt98}, while
using the 24\,$\mu$m emission, we find that
SFR=0.006\,M$_\odot$yr$^{-1}$ \citep{Calzetti05, Wu05}. This is
probably because the dust content in I\,Zw\,18 is so low while the above
relations have been calibrated for sources of high optical depth, where 
virtually all of the UV radiation is converted to infrared luminosity. A
simple calculation using the reddening curve of the Small Magellanic
Cloud assuming an A$_{\rm V}$=0.2 mag suggests that a significant
amount of UV light has leaked out without being absorbed by the dust,
thus the lower SFRs estimated from the infrared are not unexpected.
The readers should be aware of these complications when applying
the canonical infrared relations for estimating the SFR in environments
with low dust optical depth.  If we were to assume that L$_{\rm IR}$
accounts for the bolometric luminosity of the obscured populations
while L$_{\rm UV}$ \citep{Kinney93} represents the contribution of the
unobscured stars, and use the eq.1 of \citet{Bell05}\footnote
{SFR=9.8$\times$10$^{-11}$~(L$_{\rm IR}$+2.2~L$_{\rm UV}$)}, we find a
SFR of 0.02\,M$_\odot$yr$^{-1}$, more consistent with the SFRs derived
using the H$\alpha$ or radio luminosities.

\begin{deluxetable}{llc}
\tabletypesize{\scriptsize}
\setlength{\tabcolsep}{0.02in}
\tablecaption{Star Formation Rate Estimates of I\,Zw\,18\label{tab3}}
\tablewidth{0pc}
\tablehead{ 
  \colhead{SFR indicator}  & \colhead{SFR} & \colhead{Reference}\tablenotemark{a}\\
  \colhead{}  & \colhead{(M$_\odot$yr$^{-1}$)} & \colhead{}
 }
\startdata
L$_{\rm H\alpha}$\tablenotemark{b}           & 0.05  & (1) \\
L$_{\rm H\alpha}$\tablenotemark{c}           & 0.03  & (2) \\
L$_{\rm TIR}$                               & 0.003 & (3) \\
L$_{24\mu m}$                               & 0.006 & (4) \\
L$_{IR+UV}$                                & 0.02  &  (5) \\
L$_{\rm thermal}$                            & 0.1   & (6) \\
L$_{\rm non-thermal}$                         & 0.03  & (6) \\

\enddata
\tablenotetext{a} {References: 1) \citet{Kennicutt94}, 2) \citet{Lee02} 3) \citet{Kennicutt98} 4) \citet{Wu05} 
  5) \citet{Bell05} 6) \citet{Condon92}.}
\tablenotetext{b} {Not corrected for low-metallicity effects.}
\tablenotetext{c} {After correcting the low-metallicity effects.}

\end{deluxetable}

\section{Conclusions}

We have explored the mid-IR and far-IR properties of the archetype BCD
I\,Zw\,18 based on Spitzer data:

1.Using the low-resolution modules of the IRS, we have acquired the
deepest mid-IR spectrum of this galaxy obtained so far. No PAH
emission is found which confirms the absence of PAHs in low
metallicity systems.  However, the mid- to far-IR spectral slope of
I\,Zw\,18 is surprisingly similar to NGC7714, a typical starburst
galaxy with half solar metallicity. This, especially the MIPS
70\,$\mu$m detection, would suggest the presence of a significant
amount of cold dust in I\,Zw\,18. 

2.Variations in the morphology of the galaxy from 3.6 to 24$\mu$m
imaging imply that more dust emission is present in its SE component than
in the NW one. This agrees well with the results of \citet{Cannon02}.

3. The mid-IR fine-structure lines identified in the high-resolution
spectrum of I\,Zw\,18 imply a neon and sulfur abundance of 1/23 and 1/35
Z$_{\odot}$ respectively, consistent with the optically derived oxygen 
abundance of 1/30\,Z$_\odot$.

4. Estimates of the star formation rates calculated from different
indicators show considerable scatter. L$_ {\rm IR}$ and L$_{\rm 24 \mu
  m}$ give lower SFRs when compared with results using H$\alpha$ or
L$_{\rm 1.4GHz}$, probably because the low dust content in this galaxy
can only convert a small fraction of the UV radiation emitted by
stars into L$_{\rm IR}$. This should be considered when interpreting star formation
rates derived for high-redshift low metallicity galaxies.

\acknowledgments 

We thank Chad Engelbracht for graciously providing 
the IRAC and MIPS photometric measurements for our source before 
publishing the data. We also thank Rob Kennicutt for insightful 
discussions. We thank the anonymous referee, whose careful reading and 
detailed comments greatly improved this manuscript. 
Support for this work was provided by NASA 
through Contract Number 1257184 issued by JPL/Caltech.

\clearpage

\end{document}